\documentclass[epj]{svjour}
\usepackage{graphics}
\usepackage{graphicx}
\usepackage{cite}
\usepackage{amsmath}

\newcommand{\om}{\omega}

\newcommand{\rmi}[0]{\mathrm{i}}
\newcommand{\im}[0]{\mathrm{i}}
\newcommand{\bea}{\begin{eqnarray}}
  \newcommand{\beq}{\begin{equation}}
    \newcommand{\eea}{\end{eqnarray}}
  \newcommand{\eeq}{\end{equation}}

\begin{document}
\title{Fano resonances in scattering: an alternative perspective}
% \subtitle{Do you have a subtitle?\\ If so, write it here}
\author{Lukas Schwarz\inst{1} \and Holger Cartarius\inst{1} \and G\"unter Wunner\inst{1} \and Walter Dieter Heiss\inst{2,3} \and J\"org Main\inst{1}}
\institute{Institut f\"ur Theoretische Physik, Universit\"at Stuttgart,
  Pfaffenwaldring 57, 70\,569 Stuttgart, Germany\and
  Department of Physics, University of Stellenbosch,
  7602 Matieland, South Africa\and 
  National Institute for Theoretical Physics (NITheP), Western Cape,\
  South Africa}

\date{Received: date / Revised version: date}
\abstract{In a previous paper it has been shown that the interference of the first and
  second order pole of the Green's function at an exceptional point,
  as well as the interference of the first order poles 
  in the vicinity of the exceptional point, 
  gives rise to asymmetric scattering cross section profiles. 
  In the present paper we demonstrate that these
  line profiles 
  are indeed well described by the Beutler-Fano formula, and thus are
  genuine Fano resonances. 
  Also further away from the exceptional points excellent
  agreement can be found by introducing energy dependent Fano parameters.
  \PACS{03.65.Nk, 46.40.Ff, 03.65.Vf, 31.15.-p   
  } % end of PACS codes
} %end of abstract
\maketitle

\section{\label{sec:intro} Introduction}

Fano resonances were first discovered in atomic phys\-ics \cite{Beutler1935,Fano1935}.
%
%in particular in the photoionisation and scattering of atoms 
%\cite{Fano61,Fano1968,Fano1986}. 
In cross sections in nuclear physics they also appear and are
called Feshbach resonances \cite{Feshbach58,Blatt1979}.  
The characteristic features of these resonances are asymmetric line shapes, caused by the interaction
between a discrete state and a continuum of states, as described by the
formula derived by Fano in his landmark work in 1961 \cite{Fano61}. 
Later Fano himself developed
the subject further in numerous papers, review articles \cite{Fano1968}, 
and a book \cite{Fano1986}. 

To date Fano resonances have been observed in a wide array of different 
subfields of physics. They have been found, to give just a
few examples, in mesoscopic condensed matter systems \cite{Rau04}, 
in light scattering by finite obstacles \cite{Tribelsky2008}, in nanoscale 
structures \cite{Miroshnichenko10}, in wave transmission through resonant
scattering structures\cite{Rotter2010}, in broadband at\-to-\-sec\-ond-pulsed
XUV light experiments \cite{Ott2013,Lin2013}, and in coupled
plasmonic systems \cite{Lukyanchuk2010,Zhao2012,Lovera2013,Butet2014}. 

In this paper we take a different theoretical perspective on Fano resonances.
 Similar to the well-established description of an isolated 
resonance by a pole in the complex plane
of the scattering function, we here propose an understanding of Fano resonances by specific singularities denoted as
exceptional points (EPs) \cite{Kato66}.

There is a host of literature about exceptional points  \cite{Moiseyev2011a,Heiss12,Heiss14} and references quoted therein, see also, e.~g., 
\cite{Rotter2009b,Gutoehrlein13,Am-Shallem15}.
In our context the important point is the occurrence of an additional pole of second order in the scattering function at the EP.
As a consequence, the cross section (the modulus squared of the scattering function) is then characterised by the all important
interference of the poles of first and second order. This has been hinted at in a previous paper \cite{Heiss14}. The present paper
implements this claim with substantial numerical evidence.

We demonstrate that both at and in the vicinity of the
exceptional point the cross sections indeed give rise to Fano profiles locally,
and therefore can be interpreted as {\em genuine} Fano resonances. Moreover,
the cross sections can globally be described by the Beutler-Fano formula, if
energy-dependent Fano parameters are introduced. As an aside we demonstrate that
specific pa\-ra\-metrisations traditionally used \cite{Rotter2003} are bound to fail at the EP,
since the two peaks of the cross section cannot be associated with individual resonances.

\section{The model}

For the reader's convenience we briefly 
review the model introduced in \cite{Heiss14}. The model is inspired by
work of Joe et al. \cite{Joe06}, who had pointed out an  
analogy
between quantum interference in Fano resonances and classical resonances
in coupled oscillators (see also \cite{Riffe2011,Satpathy2012}).
The system consists of two one-di\-men\-sional harmonic oscillators with unperturbed eigenfrequencies $\om_1$, $\om_2$ and damping constants $k_1$, $k_2$, coupled by a spring
with spring constant $f$ and damping constant $g$, and periodically driven 
by an external force with frequency $\om$. We note that in the literature 
the model is used to describe the plasmonic
response of coupled plasmonic structures (cf. \cite{Lovera2013,Butet2014}).

After setting up the equations
of motion in phase space, 
$
(\vec{\dot p}, \vec{{\dot q}} )^{\mathrm T} = {\cal M} ({\vec p}, {\vec q})^{\mathrm T} + {\vec c} \exp(\rmi \om t) \,, 
$
resonant solutions with real frequency
$\om$ are found by the roots of the 
characteristic polynomial 
of the matrix $\cal M$.
{Additional singularites, called}
exceptional points, occur for complex values of $\om$ where in addition the
first derivative of the characteristic polynomial vanishes.  
This leads to 4 real-valued
equations for the 8 real parameters
% \beq
$\omega_1, \omega_2,k_1, k_2, g, f, \Re(\omega),$ $\Im(\omega)$.
% \eeq
If we provide specific values for the eigenfrequencies and damping constants,
we are left with 4 equations for the 4 unknowns $g, f, \Re(\omega), \Im(\omega)$,
which in general have to be solved numerically.
%In doing so the
%physical constraints on the spring and coupling constants $f >0$, $g > 0$,
%and the frequency $\Re(\omega) > 0$ and width  $\Im(\omega) < 0$ have to
%be imposed.
However, in the special case that  the individual {oscillators are undamped}, $k_1 = k_2 = 0$, 
as used in the following, the exceptional points
can be given in analytical form:
  \begin{align*}
    \Re[\omega_{\mathrm{EP}}] &= \frac{1}{2\sqrt{2}} \sqrt{\frac{\left(3 \omega_{1}^2+\omega_{2}^2\right) \left(\omega_{1}^2+3 \omega_{2}^2\right)}{\omega_{1}^2+\omega_{2}^2}} \;, \\ 
    \Im[\omega_{\mathrm{EP}}] &= \frac{-1}{2\sqrt{2}} \frac{|\omega_{1}^2-\omega_{2}^2|}{ \sqrt{\omega_{1}^2+ \omega_{2}^2}} \;,  
     \end{align*}
     \begin{align*}
    g_{\mathrm{EP}} &= - \Im[\omega_{\mathrm{EP}}] = \frac{1}{2\sqrt{2}} \frac{|\omega_{1}^2-\omega_{2}^2|}{ \sqrt{\omega_{1}^2+ \omega_{2}^2}} \;,  \\
    f_{\mathrm{EP}} &= \frac{\left(\omega_{1}^2-\omega_{2}^2\right)^2}{4 \left(\omega_{1}^2+\omega_{2}^2\right)} \;.
  \end{align*}
  
In close vicinity of an exceptional
point the higher dimensional eigenvalue problem can be reduced to an effective
two-channel scattering problem.
The advantage of the reduction is the 
analytic availability of the Green's function and thus the scattering 
amplitudes both near and at the exceptional point.
>From the Green's function the $T$ matrix 
can be calculated, and from it the  cross sections  $|T_{11}|^2$, $|T_{22}|^2$.
For details we refer the reader to \cite{Heiss14}.

\section{Fano profiles}
\subsection{Fits with one Fano profile}
{To prove that the asymmetric line profiles found in \cite{Heiss14} around exceptional points are actual Fano resonances, } 
  we {first} recall the Beutler-Fano formula (see e.g. {\bf{\cite{Fano1986}}})
\begin{equation}
  \sigma(\epsilon) = \frac{(\epsilon + q)^2}{\epsilon^2 + 1}.
  \label{Fano_1}
\end{equation}
It quite generally describes line profiles {for a physical process where 
  a continuum state interacts with a bound state embedded in the continuum.
  In (\ref{Fano_1})
  \begin{equation}
    \epsilon = \frac{E-E_{\rm R}}{\Gamma/2} 
  \end{equation}
  is the reduced energy which measures
  the energy relative to the position of the resonance $E_{\rm R}$ in units
  of the half-width $\Gamma/2$  
  being the width of the quasi-bound state in the continuum
  (note that the interaction endows the quasi-bound state with a width).}
The parameter $q$
determines the shape of the resonance and depends on the ratio 
of the transition
matrix elements linking the initial state
to the discrete and continuum parts of the final state.

\begin{figure}[tb]
  \begin{center}
    % \vspace*{-3mm}
    \includegraphics[width=0.45\textwidth]{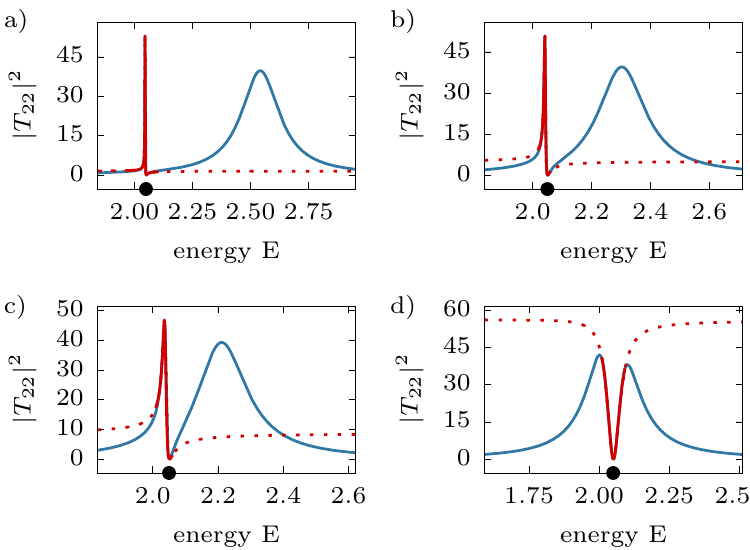}
  \end{center}
  \caption{Fano fits (red) of the cross sections at and near the exceptional point.
The parameters are $\om_1 = 2.00$ and $\om_2 = 2.10$, $k_1 = k_2 =0$, the 
    exceptional point appears at $f_{\rm EP} = 0.005$, $g_{\rm EP}= 0.0499$, and $\om_{\rm EP} = 2.0500 - 0.04999 \im$.
    (a) $f = f_{\rm EP}+1.0$, (b) $f = f_{\rm EP}+0.5$, (c) $f = f_{\rm EP}+0.3$, (d)
    $f = f_{\rm EP}$.
  }
  \label{fig_1}
\end{figure}

In the following we assume no
damping of the individual oscillators and choose
$\om_1 = 2.00$ and $\om_2 = 2.10$. The exceptional point then appears
at $f_{\rm EP} = 0.005$, $g_{\rm EP}= 0.0499$, and $\om_{\rm EP} = 2.0500 - 0.04999 \im$. 
In Fig.~{\ref{fig_1}} we {illustrate} Fano fits of the cross sections at and near the exceptional point.
The fitting parameters are $E_{\rm R}$, $\Gamma$, and $q$, in addition we have a scaling factor in front of eq. (\ref{Fano_1}) to scale the height 
of the resonance and a global shift of the cross section to account for a 
background. Numerical values of the fitting parameters are given in Table~\ref{table_1}.

It is evident from Fig.~{\ref{fig_1}} that both at and further away from the exceptional 
point {the Beutler-Fano formula locally describe the asymmetric line profiles perfectly. 
  It thus confirms that they are Fano resonances.}
\begin{table}[tb]
  \caption{ Resonance energies $E_{\mathrm{R}}$, widths $\Gamma$, and asymmetry parameter $q$ of the Fano profiles around the exceptional point shown in 
    Fig.~{\ref{fig_2}}.}
  \label{fitparameters1}
  \begin{center}
    \begin{tabular}{llll}
      \hline\noalign{\smallskip}
      & $E_{\mathrm{R}}$ & $\Gamma$ & $q$\\
      \noalign{\smallskip}\hline\noalign{\smallskip}
      a) & 2.04628 & 0.00078 & -6.43165\\
      b) & 2.04242 & 0.00289 & -3.03946\\
      c) & 2.03899 & 0.00623 & -2.09385\\
      d) & 2.04917 & 0.02755 & -0.06180\\
      \noalign{\smallskip}\hline
    \end{tabular}
  \end{center}
  \label{table_1}
\end{table}

\subsection{Fits with two Fano parameters}

\begin{figure}[tb]
  \begin{center}
    \includegraphics[width=0.45\textwidth]{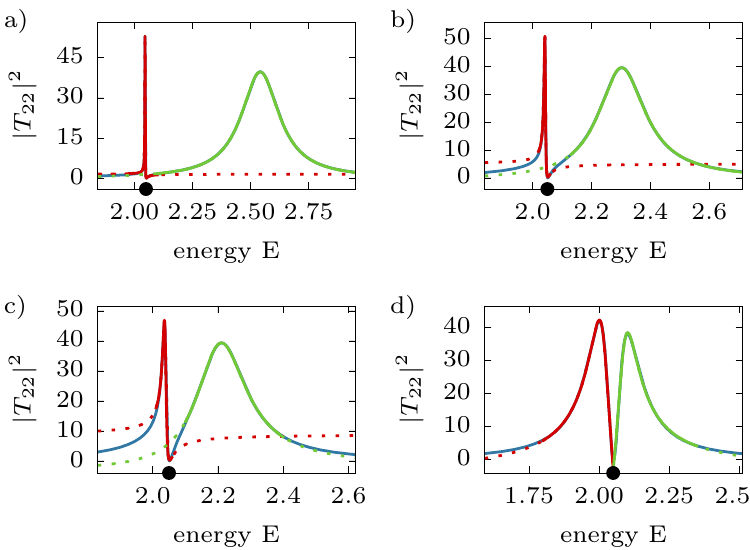}
  \end{center}
  \caption{Fano fits  of the cross sections shown in Fig~{\ref{fig_1}} with
    two sets of Fano parameters. 
  }
  \label{fig_2}
\end{figure}

If one allows for {\it two} sets of energy dependent Fano parameters 
the two peaks in the cross sections and the asymmetric line profile 
can indeed be modelled over a far wider range of the energy. Examples
are shown in Fig.~{\ref{fig_2}}, the fitting parameters are listed in
Table~\ref{table_2}. It can be seen that further away from the
exceptional point the profiles are locally well described by two Fano forms.
This, however, is not surprising since by their very nature each of the separate resonances, 
one with a larger and one with a smaller width, should be describable
by Fano profiles, with different parameters. {It is remarkable to see that 
  the peaks retain Fano profiles
  even closer to and in particular at the exceptional point, where the
  interference terms become increasingly important.}

\begin{table}[tb]
  \caption{\label{fitparameters2} Resonance energies, widths, and asymmetry parameters of the two Fano profiles shown in 
    Fig.~{\ref{fig_2}}.}
  \begin{center}
    \begin{tabular}{lllllll}
      \hline\noalign{\smallskip}
      & $E_{\mathrm{1}}$ & $\Gamma_{1}$ & $q_{1}$ & $E_{\mathrm{2}}$ & $\Gamma_{2}$ 
      & $q_{2}$\\
      \noalign{\smallskip}\hline\noalign{\smallskip}
      a) & 2.0463 & 0.0008 & -6.4770 & 2.5416 & 0.0995 & 299.0411\\
      b) & 2.0424 & 0.0029 & -3.1026 & 2.2997 & 0.0969 & 36.6894\\
      c) & 2.0390 & 0.0062 & -2.1014 & 2.2028 & 0.0934 & 12.1206\\
      d) & 2.0244 & 0.0438 & -1.8446 & 2.0785 & 0.0440 & 1.8728\\ 
      \noalign{\smallskip}\hline
    \end{tabular}
  \end{center}
  \label{table_2}
\end{table}
\subsection{Fit with energy dependent Fano parameters}
The idea of energy dependent Fano parameters has been discussed also
in the context of $q$ reversals in Rydberg series and multiphotoionisation
\cite{Connerade1987a,Connerade1987b}. Here we follow the formulation
put 
forward by Magunov et al. \cite{Rotter2003} in their study of overlapping
resonances. Starting from the $S$ matrix with  two isolated resonances
$\tilde{E}_k = {E}_k - \rmi {\Gamma}_k/2$ ($k=1,2$), and a
smooth reaction background described by a phase shift $2 \delta$,
they show that the cross section 
can be brought into the Beutler-Fano form
\begin{equation}
  \sigma(E) = 4 \sin^2 \eta(E)\frac{(\varepsilon_1+q(E))^2} {\varepsilon_1^2 +1 },
  \label{energydependent}
\end{equation}
with
\begin{equation*}
  \eta(E) = \delta - {\rm arccot} (\varepsilon_2), \quad \varepsilon_k = 2(E-{E}_k)/{\Gamma}_k,   \nonumber 
\end{equation*}
and the {\it energy dependent} Fano parameter
\begin{equation*}
  q(E) = -\cot \eta(E) .
\end{equation*}

Apart from the scale factor for the height of the cross
section,  in the extended ansatz (\ref{energydependent}) the 
fitting parameters are $E_{1}$, $\Gamma_1$,  $E_2$, $\Gamma_2$ and the
phase $\delta$. 

In all our fits the numerical value of $\delta$ 
turned out to be close to $\pi$ up to the first five digits. Since
$\delta$ enters into the $S$ matrix by the factor $\exp({\rmi 2 \delta})$, this 
implies that in our model there is no background phase, and we could
have chosen $\delta = 0$ from the outset. In this  
case  (\ref{energydependent}) simplifies to
\begin{equation}
  \sigma(E) = \frac{1}{(\epsilon_2^2+1)} \frac{(\epsilon_1 + \epsilon_2)^2}{(\epsilon_1^2+1)} 
\;.
  \label{sigma_simplified}
\end{equation}
The comparison with (\ref{Fano_1}) then shows that $\epsilon_2$ formally assumes the role of an energy dependent asymmetry parameter $q(E)$ for resonance 1, and {\it vice versa}.

\begin{table}[tb]
  \caption{Fitting parameters for the cross sections shown in Fig.~\ref{fig_3}
    calculated using eq. (\ref{sigma_simplified}).}
  \centering
  \begin{tabular}{lllll}
    \hline\noalign{\smallskip}  
    ~ & $E_\mathrm{1}$ & $\Gamma_1$ & $E_\mathrm{2}$ & $\Gamma_2$ \\
    \noalign{\smallskip}\hline\noalign{\smallskip}      
    a)  & 2.04970 & 0.10792 & 2.04990 & -0.09239 \\
    b)  & 2.49559 & 1.04493 & 2.50780 & -0.91573 \\
    c)  & 3.47188 & 3.20257 & 3.62857 & -2.28382 \\
    d)  & 2.03912 & 0.01471 & 2.20748 & 0.18673 \\
    e)  & 2.04252 & 0.00680 & 2.30187 & 0.19218 \\
    f)  & 2.04626 & 0.00196 & 2.54172 & 0.19923 \\
    \hline\noalign{\smallskip}        
  \end{tabular}
  \label{table_3}
\end{table}
\begin{figure}[tb]
  \begin{center}
    \includegraphics[width=0.45\textwidth]{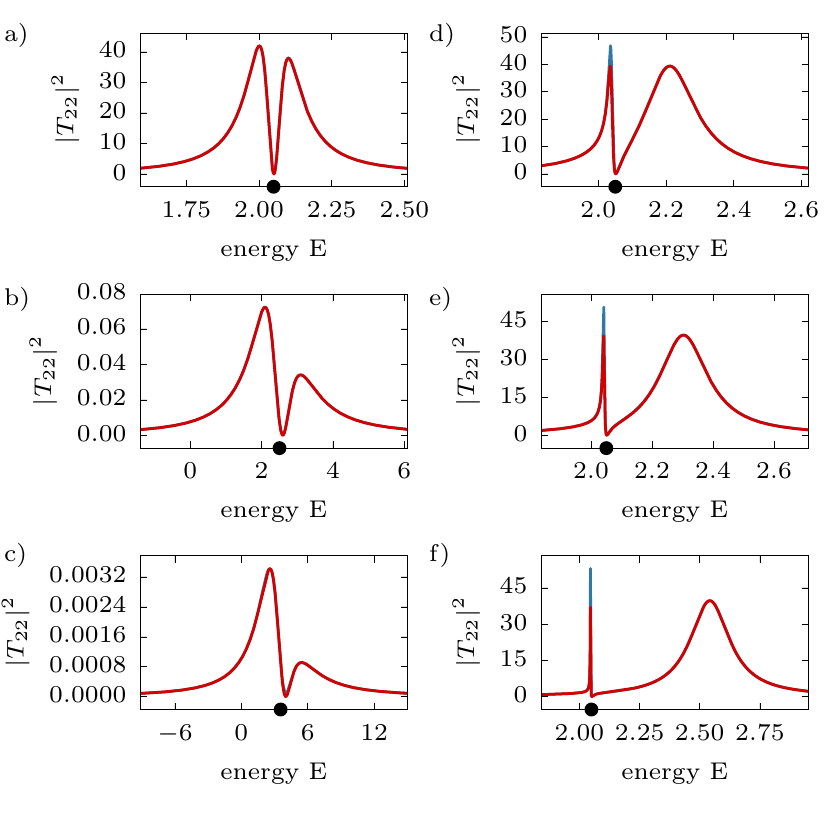}
  \end{center}
  \caption{Fano fits of cross sections at and in the vicinity of exceptional points using eq. (\ref{sigma_simplified}). The blue curves are the calculated
    cross sections, the red curves denote the fits.\hfill\break
    The left-hand column shows the results for $\omega_1 = 2.00$ 
    and $\omega_2 = 2.10, 3.00, 5.00$ (from top to bottom) at the respective 
    exceptional point (black dots), while in the right-hand column
    $\omega_1 = 2.00$, $\omega_2 = 2.10$, and the distance from the
    exceptional point is successively increased, $\Delta f = 0.3, 0.5, 1$ (from
    top to bottom).
  }
  \label{fig_3}
\end{figure}

Examples for fits of the cross sections with energy dependent asymmetry parameters at and in the vicinity of exceptional points  are shown in Fig.~{\ref{fig_3}}. 
The agreement between the calculated cross section and the
energy dependent Fano fit is found to be perfect over the complete energy range. 

The fitting parameters required for this excellent agreement are listed in
Tab.~\ref{table_3}. For sufficient distance from the EP (rows d-f), where the interference term
plays a minor role, this is to be expected as we encounter essentially two independent
and well separated resonances associated with poles the positions of which are given in the table.
In contrast, at the EP (rows a-c) the two peaks cannot be {identified} as two separated resonances; in fact, here
the interference term is crucial to produce the two peaks (see \cite{Heiss14}). It therefore comes as no
surprise that the best fits produce results - negative widths - that can no longer be interpreted as
physical resonances. {This clearly indicates that the treatment \cite{Rotter2003} by energy dependent Fano parameters fails at an EP.} 
In fact, the double pole invoked by the EP is different in character from the double pole
as considered in \cite{Rotter2003}: the two-dimensional coefficient matrix of the double pole at the EP does not have full rank
but rank unity (see \cite{Heiss12})
meaning that the entries are correlated in a particular way. In this context, we also note that our scattering
matrices are not unitary as the underlying Hamiltonian is not hermitian.

\section{Summary}
Starting from the fact that in scattering systems exceptional points give rise to a second order pole in the Green's function, 
we have investigated the effect on the shape of the scattering cross section as one approaches the exceptional point.
In this paper we have demonstrated that the asymmetric line profiles in the 
neighbourhood of the exceptional point are to be interpreted as Fano resonances.
Moreover, by allowing for energy dependent Fano para\-meters the cross sections
can also globally be described as actual Fano profiles.

We feel that by this analysis we have given a new and deeper understanding of Fano resonances.
On the one side there is the mathematical mechanism of two coalescing eigenvalues - an EP -
on the other the physical origin of two interacting near resonances where the one
can be a single particle resonance and the other a bound state in the continuum that,
owing to the interaction, has acquired itself a width. This is the typical
Fano-Feshbach scenario. Whether or not all resonances that can be fitted by the traditional
Fano-Feshbach procedure have their origin in an EP cannot be shown. In fact, the particular
type and relative strength of the interaction must be taken into account. Yet we believe
that the mechanism proposed here should be representative quite generally.

It should be stressed that, in spite of 
the simplicity of the underlying classical problem, the generic behaviour of
line profiles studied in this paper near exceptional points applies
to any physical two-channel scattering or transmission problem. 
Therefore we did establish the direct link between the appearance of asymmetric 
Fano profiles and the occurrence of exceptional points in parameter space. 

\section*{Acknowledgement}
WDH and GW gratefully acknowledge the support from the National Institute for Theoretical Physics (NITheP), 
Western Cape, South Africa. GW expresses his gratitude to the Department of Physics of the University of
Stellenbosch where this work was started and completed.

\bibliographystyle{unsrt}
%\bibliography{literature}
%\end{document}

\end{document}